\documentclass[pra,reprint]{revtex4-1}

\usepackage{graphicx}
\usepackage{verbatim,pgf}
\usepackage{bbold}
\usepackage{amsmath}

\begin{document}

\title{Sub-fidelity as a measure of irreversibility of decoherence}
\author{Filippo M. Miatto$^{1,2}$, Kevin Pich\'e$^{1,3}$, Thomas Brougham$^{3}$, and Robert W.~Boyd$^{1,3,4}$}
\affiliation{$^1$Dept. of Physics and Max Planck Centre for Extreme and Quantum Electronics, University of Ottawa, Ottawa, Canada}
\affiliation{$^2$Institute for Quantum Computing, University of Waterloo, Waterloo, Canada}
\affiliation{$^3$School of Physics and Astronomy, University of Glasgow, Glasgow (UK)}
\affiliation{$^4$Institute of Optics, University of Rochester, Rochester, USA}
\date{\today}

\begin{abstract}
%The properties of quantum systems cannot be always deemed as objective as those of a classical system, because quantum correlations make subsystems ``steerable'', i.e. we can choose a preferred basis in the Hilbert space of a system by acting on a correlated one. 
Quantum correlations between two systems can be impaired by the presence of an environment, as it can make systems decohere. We wish to evaluate how much coherence has been irreversibly lost. To do so, we study two qubits which leak correlations into an arbitrarily large environment and we find that our ability to steer one of them to the most coherent set of states possible by acting on the other is given by the sub-fidelity of two conditional states of the environment. Sub-fidelity is a quantity that was already known, but that until now lacked an operational interpretation. Furthermore, we conjecture the existence of a family of measures, of which fidelity and sub-fidelity are the first two, which measure the irreversibility of decoherence as we gain access to more and more degrees of freedom of the enviroment.
\end{abstract}
\maketitle

\section{Introduction}
%Complementarity is one of the jewels of quantum mechanics. It was first introduced by Bohr \cite{Bohr1}, as a consequence of the uncertainty principle. However, it took several decades until Scully and his colleagues established that its origin is really quantum correlations \cite{Scully1991,Wiseman1995,Durr1998}, and in ultimate analysis it can be ascribed to the entropic uncertainty principle \cite{coles2014equivalence}. The principle of complementarity gained its modern form through the works of several authors \cite{WZ1972,GreenbergerYasin,Scully1989,Sanders1989,Durr1998}. In particular, Englert gave a lucid exposition of the connection between complementarity and the working principles of a two-way interferometer \cite{Englert1996}.

Quantum correlations can be stronger than their classical counterparts because they enable effects that are classically unobserved. A great example of such departure from classicality is the EPR paradox, where correlations seem to allow for measurements in complementary bases. It is precisely such freedom to choose a preferred basis in the Hilbert space of a quantum system by interacting with a correlated one that is behind quantum steering. 
``Steering'' is a term coined by Schr\"odinger: ``It is rather discomforting that the theory should allow a system to be steered or piloted into one or the other type of state at the experimenter's mercy in spite of his having no access to it.'' \cite{schrodinger1935discussion}. The simplest example of steering involves two correlated qubits ($A$ and $B$) and two parties (Alice and Bob), where Alice steers Bob's qubit by acting locally on hers \cite{wiseman2007steering,piani2015necessary}. In this work we consider a particular example of steering, where Alice wants to steer Bob's qubit $B$ to the most coherent set of states possible. This type of steering can also be described as quantum erasure \cite{miatto2015recovering}.

We can understand the concept of coherence of a state through our familiarity with the Bloch sphere and by an analogy with Mach-Zender interferometers \cite{Scully1982,GreenbergerYasin,Scully1989,Kwiat1992,Herzog1995,Englert1996,Durr1998}. Given a state with Bloch vector $\mathbf{v}=(x,y,z)$, its coherence is the distance from the North-South axis, i.e. $\mathcal{C}(\mathbf{v})=\sqrt{x^2+y^2}$, which can be calculated from a qubit density matrix as twice the absolute value of one of the off-diagonal elements: $\mathcal C=2|\langle 0|\hat\psi|1\rangle|$, where $\hat\psi=\frac{1}{2}(\mathbb{\hat1}+\mathbf{v}\cdot\mathbf{\sigma})$ is the density operator corresponding to the Bloch vector $\mathbf{v}$ and $\sigma={(\sigma_x,\sigma_y,\sigma_z)}$ is the vector of Pauli operators. It is clear that coherence depends on our choice of North and South poles, which are also called the ``alternatives'', implying a measurement of $\sigma_z$. We can give a useful example by describing a photon travelling through a balanced Mach-Zender interferometer: the two paths constitute the two possible alternatives and so we can represent the path of the photon with a qubit. As one varies the phase of one arm with respect to the other, the Bloch vector rotates around the $z$ axis in a circular motion of radius $\sqrt{x^2+y^2}$. Such radius is precisely the visibility of the fringes at the output of the interferometer. Now suppose that the initial beam splitter is a polarizing one (PBS), so that the path of the photon can be learned at any moment by a measurement of polarization. In this situation we have two qubits (one is the path, the other is the polarization) which are maximally correlated, so the state of the path qubit alone is maximally mixed. A maximally mixed state has Bloch vector $\mathbf{v}=(0,0,0)$, and correspondingly zero coherence, which is why we would not observe any fringes at the output of the interferometer. Quantum erasure would instruct us to erase the path information that is stored in the polarization qubit by using (for instance) a  diagonally-oriented PBS to close the interferometer. Such procedure is actually an instance of steering: the state of the photon in the interferometer is $\frac{1}{\sqrt{2}}[|0,H\rangle+e^{i\phi}|1,V\rangle]$, where $|0\rangle$ and $|1\rangle$ indicate the paths. This state can be re-written as $\frac{1}{2}[|D\rangle(|0\rangle+e^{i\phi}|1\rangle)+|A\rangle(|0\rangle-e^{i\phi}|1\rangle)]$, i.e. if we could project the polarization qubit in the $|A\rangle/|D\rangle$ basis, we would steer the path qubit towards the equator of the Bloch sphere, i.e. to two states displaying perfect coherence. We cannot predict which of the two ports will output the photon, but we do know that one of them will, and in either case, the states exhibit maximal coherence. This is the essence of quantum erasure.

Such procedure can be impaired by the presence of a third system, which learns about the state of one of the qubits \cite{Englert2000,leach2014duality,bolduc2014fair}. The reason why we may encounter difficulties in steering a qubit towards a highly coherent set of states is because of the monogamy of quantum correlations: as the environment becomes more correlated with the two qubits, the correlations between them must decrease. Consequently, their steerability also suffers. Can we make this intuition more precise? Specifically, can we quantify the fraction of coherence that leaked into the environment in an irreversible way? To do so we need to evaluate what the environment knows.

\section{derivation of the main result}
To set the scene, let's assign one qubit to Alice ($A$) and the other one to Bob ($B$) and let's start with a simpler task: we want to find the coherence of Bob's qubit $B$ when Alice does not act upon her qubit or any of the systems that are correlated to $B$, i.e. the ``no-steering'' case. To do so we write the state of $B$ and of its environment (which in this case must include Alice's qubit) as
\begin{align}
|\psi\rangle=\alpha|0,e_0\rangle+\beta|1,e_1\rangle
\end{align}
which is pure because we took into account all the necessary environment. Then, by definition, the absolute value of the off-diagonal term in the reduced density matrix of Bob's qubit $B$ yields the coherence, which is
\begin{align}
\label{simple}
\mathcal C_1&=2|\alpha\beta^*\langle e_0|e_1\rangle|\\
&=2\sqrt{p_0p_1F(\hat\psi_E^{(0)},\hat\psi_E^{(1)})}
\label{fidelity}
\end{align}
where $p_0=|\alpha|^2$ and $p_1=|\beta|^2$ are the probabilities of observing the two alternatives of $B$, $\hat\psi_E^{(k)}=|e_k\rangle\langle e_k|$ is the state of the environment conditional on the state of $B$ and $F(x,y)=\mathrm{Tr}(x^\dagger y)$ is the Hilbert-Schmidt inner product, which evaluates the overlap and which for pure states coincides with the fidelity \cite{nielsen2010quantum}. Eq.~\eqref{fidelity} bears a remarkable similarity with the main result in Eq.~\eqref{result}.

Let's now move on to the steering case. Consider the state of both qubits, $\hat\psi_{AB}=\mathrm{Tr}_E[\hat\psi_{ABE}]$, obtained by tracing over the environment. Then, let Alice perform a measurement on $A$, defined by a complete set of probability operators $\hat\pi_A^{(k)}$. Given outcome $k$, Bob's qubit is projected to the state $\hat\psi_B^{(k)}=\mathrm{Tr}_A[\hat\pi^{(k)}_A\hat\psi_{AB}]/p_k$ with probability $p_k=\mathrm{Tr}[\hat\pi^{(k)}_A\hat\psi_{AB}]$.
The coherence of Bob's state, averaged over all the outcomes of Alice's measurement, is therefore
\begin{align}
\mathcal{\overline C}&=\sum_kp_k\mathcal C(\hat \psi_B^{(k)})\\
&=\sum_k2|\langle0_B|\mathrm{Tr}_A[\hat\pi^{(k)}_A\hat\psi_{AB}]|1_B\rangle|.
\end{align}
We then maximize it over all the possible measurements that Alice can perform on her qubit \cite{Englert2000}: 
\begin{align}
\mathcal C_\mathrm{2}=\sup_{POM_A}\mathcal{\overline C}=2\mathrm{Tr}|\hat\chi_A|
\end{align}
where $\hat\chi_A=\langle0_B|\hat\psi_{AB}|1_B\rangle$. The number 2 in the subscript reminds us that Alice is steering Bob's qubit by acting on a 2-dimensional system (her qubit).
After a bit of algebra (see appendix) we find
\begin{align}
\label{result}
\mathcal C_{2}=2\sqrt{p_0p_1E(\hat \psi_{E}^{(0)},\hat \psi_{E}^{(1)})}
\end{align}
where $E(\hat \psi_{E}^{(0)},\hat \psi_{E}^{(1)})$ is the sub-fidelity of $\hat \psi_{E}^{(0)}$ and $\hat \psi_{E}^{(1)}$ (which are still the states of the environment conditioned on the alternatives of Bob's qubit). The sub-fidelity \cite{miszczak2009sub} is a lower bound of Uhlmann's fidelity and is defined as $E(x,y)=\mathrm{Tr}(xy)+\sqrt{2}\sqrt{\mathrm{Tr}(xy)^2-\mathrm{Tr}(xyxy)}$.

The method for deriving Eq.~\eqref{result} can be in principle generalized for higher dimensions, in order to obtain a family of measures $\{\mathcal C_a\}$ which we conjecture to be in the form
\begin{align}
\mathcal C_{a}=2\sqrt{p_0p_1F_a(\hat \psi_{E}^{(0)},\hat \psi_{E}^{(1)})}
\end{align}
where the functions $F_a$ involve only traces of products of $\hat \psi_{E}^{(0)}$ and $\hat \psi_{E}^{(1)}$. Such measures evaluate the distinguishability of the conditional states of the environment (i.e. how much the environment ``knows'') while taking into account the number of degrees of freedom of the system that Alice is operating on to steer Bob's qubit. Even if we do not find the explicit form of the measures $\mathcal C_a$ for $a\geq3$, we can evaluate their expectation value on ensembles of random states: given a uniformly random state $\hat\psi_{ABE}$, where the environment is $K$-dimensional we have from \cite{miatto2015recovering}
\begin{align}
\langle\mathcal C_1\rangle&=\frac{(-1)^K\pi ^{3/2}}{2 K! \Gamma \left(\frac{1}{2}-K\right)}\\
\langle\mathcal C_2\rangle&=\frac{(-1)^K\pi ^{3/2} (13-22K)}{32 K! \Gamma \left(\frac{3}{2}-K\right)}\\
\langle\mathcal C_3\rangle&=\frac{(-1)^K\pi ^{3/2} (433-936K+428K^2)}{512 K! \Gamma \left(\frac{5}{2}-K\right)}\\
\mathrm{etc}\dots\nonumber
\end{align}

It is interesting that the conditional states of the environment contain sufficient information to evaluate the amount of coherence of $B$ that was irreversibly lost. One would expect that some information on the correlation between $A$ and $B$ that pertains uniquely to the state $\hat\psi_{AB}$ were also necessary, as we are considering only local operations on $A$. However, there is sufficient information on the correlations between $A$ and $B$ in the states $\hat\psi_E^{(0)}$ and $\hat\psi_E^{(1)}$ and the sub-fidelity is the right function to evaluate it exactly. This fact supplies an operational meaning to sub-fidelity.

\section{Conclusion}
In this work we have considered how decoherence affects the steering of a qubit. Decoherence dissipates into the environment some information about the alternatives of the qubit by making more distinguishable the states of the environment that are conditional on such alternatives. This means that after decoherence, by looking at the environment we can do a better job at predicting which of the two alternatives would occur upon a measurement of the qubit. Now there is an ``element of the reality of the alternatives'' in the environment, i.e. the alternatives have become more objective, and this limits our ability to steer the qubit by acting on a correlated one. While it is to be expected that some measure of distinguishability between the conditional states of the environment would evaluate the coherence of the qubit that is still recoverable, it was unknown what such distinguishability measure should be. We found that it is given by the sub-fidelity, which for the first time gains an operational meaning. We finally conjecture a generalization to higher dimensions.

\section{Acknowledgements}
F.M.M. thanks Steve Barnett, Gerardo Adesso, Marco Piani and Karol \.Zyczkowski for helpful comments.
This work was supported by the Canada Excellence Research Chairs (CERC) Program, the Natural Sciences and Engineering Research Council of Canada (NSERC) and the UK EPSRC.

\section{Appendix}
We now provide our derivation of $\mathcal C_2$. We recall that the trace norm of a matrix $x$, corresponds to the sum of the singular values of $x$, which are the eigenvalues of the positive matrix $|x|=\sqrt{x^\dagger x}$.
Call $\alpha$ and $\beta$ the two eigenvalues of $x^\dagger x$, it then holds that $\mathrm{Tr}|\hat\chi_A|=\sqrt{\alpha}+\sqrt{\beta}$. We can express the sum of two square roots in terms of the elementary symmetric polynomials in two variables $s_1=\alpha+\beta$ and $s_2=\alpha\beta$ as
\begin{align}
\label{key}
\mathrm{Tr}|\hat\chi_A|=\sqrt{\alpha}+\sqrt{\beta}=\sqrt{s_1+2\sqrt{s_2}}.
\end{align}
The last thing to do is to express the symmetric polynomials in terms of traces, which can be done elegantly via Newton's identities:
\begin{subequations}
\label{newton}
\begin{align}
s_1&=\mathrm{Tr}(y)\\
2s_2&=\mathrm{Tr}(y)^2-\mathrm{Tr}(y^2)\\
6s_3&=\mathrm{Tr}(y)^3-3\mathrm{Tr}(y)\mathrm{Tr}(y^2)+2\mathrm{Tr}(y^3)\\
\mathrm{etc\ }&\dots\nonumber
\end{align}
\end{subequations}
where for us $y=\hat\chi_A^\dagger \hat\chi_A$.
So we expand $\hat\chi_A$ in its most general form:
\begin{align}
\frac{1}{\sqrt{p_0p_1}}\left( 
\begin{array}{c@{}c}
 \sqrt{r_0s_0}\langle e_{10}|e_{00}\rangle & e^{i\theta'}\sqrt{r_0s_1}\langle e_{11}|e_{00}\rangle\\
 e^{-i\theta}\sqrt{r_1s_0}\langle e_{10}|e_{01}\rangle & e^{-i(\theta-\theta')}\sqrt{r_1s_1}\langle e_{11}|e_{01}\rangle\\
\end{array}\right),
\end{align}
where $|e_{ba}\rangle$ are the states of $E$ conditioned on the alternatives of qubits $B$ and $A$. The positive numbers $r_a$ and $s_a$ are the relative probabilities of $|e_{0a}\rangle$ and $|e_{1a}\rangle$, respectively. $\theta$ and $\theta'$ are the phases of the states of $A$ conditioned on the alternatives of $B$. For simplicity, we will rewrite this as
\begin{align}
\hat\chi_A=\begin{pmatrix}
 a & c\\
 d & b\\
\end{pmatrix}.
\end{align}
Plugging this into Eq.~\eqref{key} gives us
\begin{align}
\mathrm{Tr}|\hat\chi_A|^2=|a|^2+|b|^2+|c|^2+|d|^2+2|ab-cd|.
\end{align}
Expanding $|a|^2+|b|^2+|c|^2+|d|^2$ we obtain
\begin{align}
\mathrm{Tr}(\tilde\psi_{E|00}\tilde\psi_{E|10})&+\mathrm{Tr}(\tilde\psi_{E|00}\tilde\psi_{E|11})+\nonumber\\\mathrm{Tr}(\tilde\psi_{E|01}\tilde\psi_{E|10})&+\mathrm{Tr}(\tilde\psi_{E|01}\tilde\psi_{E|11})\nonumber\\=\mathrm{Tr}(\hat\psi_{E}^{(0)}\hat\psi_{E}^{(1)})
\end{align}
where $\tilde\psi_{E|0a}=r_a|e_{0a}\rangle\langle e_{0a}|$ and $\tilde\psi_{E|1a}=s_a|e_{1a}\rangle\langle e_{1a}|$ are \emph{unnormalized} states. Consequently, $\hat\psi_{E}^{(b)}=\tilde\psi_{E|b0}+\tilde\psi_{E|b1}$ are the normalized states of $E$ conditioned on $B$ while ignoring (tracing away) $A$.
The final term is not as straightforward. We begin first by rewriting $|ab-cd|$ as $\sqrt{(ab-cd)(a^*b^*-c^*d^*)}$. We expand what is under the square root and then add and subtract the following term:
\begin{align}
\mathrm{Tr}(\tilde\psi_{E|00}\tilde\psi_{E|11})&\mathrm{Tr}(\tilde\psi_{E|01}\tilde\psi_{E|11})+\nonumber\\
\mathrm{Tr}(\tilde\psi_{E|00}\tilde\psi_{E|10})&\mathrm{Tr}(\tilde\psi_{E|01}\tilde\psi_{E|10}).
\end{align}
We then simplify the result with the identity $\mathrm{Tr}(XY)\mathrm{Tr}(XZ)=\mathrm{Tr}(XYXZ)$, which holds for $X$ rank-1. We finally obtain
\begin{align}
\mathrm{Tr}|\tilde\chi_A|^2&=\mathrm{Tr}(\hat\psi_{E}^{(0)}\hat\psi_{E}^{(1)})\nonumber\\+&\sqrt{2}\sqrt{[\mathrm{Tr}(\hat\psi_{E}^{(0)}\hat\psi_{E}^{(1)})]^2-\mathrm{Tr}[(\hat\psi_{E}^{(0)}\hat\psi_{E}^{(1)})^2]}\nonumber\\
&=E(\hat\psi_{E}^{(0)},\hat\psi_{E}^{(1)})
\end{align}

Following similar steps we can extend our analysis to the case $\mathrm{dim}(A)=3$, i.e. the case where one can access a three-dimensional subspace of the environment. In this case we have three singular values, and some simple algebra will tell us that
\begin{align}
\mathrm{Tr}|\tilde\chi_A|&=\sqrt{\alpha}+\sqrt{\beta}+\sqrt{\gamma}\nonumber\\
&=\sqrt{s_1+2\sqrt{s_2+2\sqrt{s_3}\,\mathrm{Tr}|\tilde\chi_A|}}
\label{dim3}
\end{align}
where now the symmetric polynomials are in three variables: $s_1=\alpha+\beta+\gamma$, $s_2=\alpha\beta+\beta\gamma+\gamma\alpha$ and $s_3=\alpha\beta\gamma$ and they still satisfy Eq.~\eqref{newton}. So one can solve Eq.~\eqref{dim3} for $\mathrm{Tr}|\tilde\chi_A|$ and find $\mathcal C_3$. It is in principle possible to extend this method to higher dimensions and find a whole family of distinguishability measures, but it becomes quickly intractable because the number of terms that are necessary grows very rapidly and the degree of the equations to solve increases.

\bibliography{Manuscript}
\bibstyle{unsrt}

\end{document}